\documentclass[superscriptaddress,aps,preprintnumbers,amsmath,showpacs,amssymb,prd,nofootinbib,reprint]{revtex4-1}

\usepackage[dvipdfmx]{graphicx}
\usepackage{amsfonts}
\usepackage[mathscr]{eucal}
\usepackage{ascmac}
\usepackage{dcolumn}% Align table columns on decimal point
\usepackage{bm}% bold math
\usepackage[colorlinks=true,linkcolor=blue,citecolor=blue]{hyperref}
\usepackage{hyperref}
\usepackage{color}

\begin{document}

%%%%%%%%%%%%%%%%%%%%%%%%%%%%%%%%%%%%%%%%%%%

%\newcommand{\gtrsim}{ \mathop{}_{\textstyle \sim}^{\textstyle >} }
%\newcommand{\lesssim}{ \mathop{}_{\textstyle \sim}^{\textstyle <} }
\newcommand{\vev}[1]{ \left\langle {#1} \right\rangle }
\newcommand{\bra}[1]{ \langle {#1} | }
\newcommand{\ket}[1]{ | {#1} \rangle }
\newcommand{\EV}{ \ {\rm eV} }
\newcommand{\KEV}{ \ {\rm keV} }
\newcommand{\MEV}{\  {\rm MeV} }
\newcommand{\GEV}{\  {\rm GeV} }
\newcommand{\TEV}{\  {\rm TeV} }
\newcommand{\1}{\mbox{1}\hspace{-0.25em}\mbox{l}}
\newcommand{\Red}[1]{{\color{red} {#1}}}

\newcommand{\lmk}{\left(}  
\newcommand{\rmk}{\right)}
\newcommand{\lkk}{\left[}  
\newcommand{\rkk}{\right]}
\newcommand{\lhk}{\left \{ }  
\newcommand{\rhk}{\right \} }
\newcommand{\del}{\partial}  
\newcommand{\la}{\left\langle} 
\newcommand{\ra}{\right\rangle}
\newcommand{\half}{\frac{1}{2}}

\newcommand{\bea}{\begin{array}}
\newcommand{\eea}{\end{array}}
\newcommand{\beq}{\begin{eqnarray}}
\newcommand{\eeq}{\end{eqnarray}}

\newcommand{\dd}{\mathrm{d}}
\newcommand{\Mpl}{M_{\rm Pl}}
\newcommand{\mg}{m_{3/2}}
\newcommand{\abs}[1]{\left\vert {#1} \right\vert}
\newcommand{\mphi}{m_{\phi}}
\newcommand{\Hz}{\ {\rm Hz}}
\newcommand{\for}{\quad \text{for }}
\newcommand{\Min}{\text{Min}}
\newcommand{\Max}{\text{Max}}
\newcommand{\Kahler}{K\"{a}hler }
\newcommand{\cphi}{\varphi}
\newcommand{\Tr}{\text{Tr}}
\newcommand{\diag}{{\rm diag}}

\newcommand{\SUf}{SU(3)_{\rm f}}
\newcommand{\Upq}{U(1)_{\rm PQ}}
\newcommand{\Zpq}{Z^{\rm PQ}_3}
\newcommand{\Cpq}{C_{\rm PQ}}
\newcommand{\ubar}{u^c}
\newcommand{\dbar}{d^c}
\newcommand{\ebar}{e^c}
\newcommand{\nubar}{\nu^c}
\newcommand{\Ndw}{N_{\rm DW}}
\newcommand{\Fpq}{F_{\rm PQ}}
\newcommand{\fpq}{v_{\rm PQ}}
\newcommand{\Br}{{\rm Br}}
\newcommand{\Lag}{\mathcal{L}}
\newcommand{\Lqcd}{\Lambda_{\rm QCD}}

%%%%%%%%%%%%%%%%%%%%%%%%%%%%%%%%%%%%%%%%%%%%%%%%%%%%%%%%%%%%%%%

\preprint{
IPMU 15-0085
}

\title{
Cosmologically~safe~QCD~axion 
as~a~present~from~extra~dimension
}

\author{
Masahiro Kawasaki
}
\affiliation{Institute for Cosmic Ray Research, 
The University of Tokyo, 
Kashiwa, Chiba 277-8582, Japan}
\affiliation{Kavli IPMU (WPI), UTIAS, 
The University of Tokyo, 
Kashiwa, Chiba 277-8583, Japan}

\author{
Masaki Yamada
}
\affiliation{Institute for Cosmic Ray Research, 
The University of Tokyo, 
Kashiwa, Chiba 277-8582, Japan}
\affiliation{Kavli IPMU (WPI), UTIAS, 
The University of Tokyo, 
Kashiwa, Chiba 277-8583, Japan}

\author{
Tsutomu T. Yanagida
}
\affiliation{Kavli IPMU (WPI), UTIAS, 
The University of Tokyo, 
Kashiwa, Chiba 277-8583, Japan}

\date{\today}

\begin{abstract} 
We propose a QCD axion model where the origin of PQ symmetry and suppression of axion isocurvature perturbations are explained by introducing an extra dimension. Each extra quark-antiquark pair lives on branes separately to suppress PQ breaking operators. The size of the extra dimension changes after inflation due to an interaction between inflaton and a bulk scalar field, which implies that the PQ symmetry can be drastically broken during inflation to suppress undesirable axion isocurvature fluctuations.

\end{abstract}

%\pacs{98.80.Cq, 95.35.+d, 12.60.Jv}

\maketitle

%%%%%%%%%%%%%%%%%%%%%%%%%%%%%%%%%%%%%%%%%%%%%%%%%%%%%%%%%%%%%%%

%%%%%%%%%%%%%%%%%%%%%%%%%%%%%%%%%%%%%%%%%%%%%%%%%%%%%%%%%%%%%%%%
\section{Introduction
\label{sec:introduction}}
%%%%%%%%%%%%%%%%%%%%%%%%%%%%%%%%%%%%%%%%%%%%%%%%%%%%%%%%%%%%%%%%

The null result of the electric dipole moment for neutron implies that 
the CP phase in the strong sector is smaller than of order $10^{-10}$~\cite{Baker:2006ts}. 
An elegant mechanism to explain the smallness of the strong CP phase 
is the PQ mechanism~\cite{Peccei:1977hh, Peccei:1977ur}. 
The strong CP phase is promoted to a dynamical field, called axion, 
which is associated with a spontaneously broken anomalous symmetry~\cite{Weinberg:1977ma}. 
The axion obtains a periodic potential through nonperturbative effects~\cite{'tHooft:1976up, 'tHooft:1976fv} 
and starts to oscillate around its minimum at the QCD phase transition. 
The strong CP phase is dynamically cancelled by the vacuum expectation value (VEV) of the axion. 
In addition, 
the axion oscillation behaves like the cold dark matter (DM) 
and can explain the observed amount of DM~\cite{Preskill:1982cy, Abbott:1982af, Dine:1982ah}. 
However, 
the PQ symmetry is an anomalous global symmetry, 
so that it is not trivial that Lagrangian respects that symmetry in a sufficiently high accuracy~\cite{
Giddings:1988cx, Coleman:1988tj, Gilbert:1989nq, Banks:2010zn}. 
In fact, in the literature they pointed out that 
the PQ symmetry may be broken by quantum gravitational effects, 
so that severe fine-tunings may be required to solve the strong CP problem by the PQ mechanism~\cite{Kamionkowski:1992mf, Ghigna:1992iv, Dobrescu:1996jp}.

In Refs.~\cite{Izawa:2002qk, Izawa:2004bi}, Izawa, Watari and one of the present authors proposed a model 
that has an accidental PQ symmetry by introducing an extra dimension.%
\footnote{
See Refs.~\cite{Holman:1992us, Barr:2001vh, Harigaya:2015soa} for other mechanisms 
to explain the origin of PQ symmetry. 
}
Each extra quark-antiquark pair lives on branes separately 
so that PQ breaking operators are suppressed exponentially by the size of the extra dimension. 
The extra quarks are charged under QCD and a bulk hypercolor 
that is assumed to be confined at an intermediate scale of order $10^{12} \GEV$. 
Its strong dynamics makes the extra quarks confined 
and breaks the accidental PQ symmetry at that scale. 
In this model, a composite NG boson plays the role of axion 
and the PQ mechanism is realized without fine-tunings~\cite{Kim:1979if}.

Although the above scenario confronts with the cosmological domain wall problem~\cite{Zeldovich:1974uw, Sikivie:1982qv}, 
it can be avoided when the PQ symmetry is broken before inflation. 
In this case, however, the axion DM predicts sizable isocurvature density perturbations~\cite{Axenides:1983hj, Seckel:1985tj, Turner:1990uz} 
due to quantum fluctuations in the axion field during inflation. 
The resulting amplitude of isocurvature perturbations is inconsistent with the observation of CMB fluctuations 
unless the energy scale of inflation $H_{\rm inf}$ is smaller than about $10^7 \GEV$. 
Such a small energy scale excludes many interesting inflation models, such as the Starobinsky model~\cite{Starobinsky:1980te} 
and chaotic inflation model~\cite{Linde:1983gd}.

In this letter, 
we investigate a stabilization mechanism of the size of extra dimension 
and propose a scenario to suppress axion isocurvature perturbations.%
\footnote{
See Refs.~\cite{Linde:1990yj, Linde:1991km, Kasuya:1996ns, Kasuya:1997td, Folkerts:2013tua, Jeong:2013xta, Kawasaki:2013iha, Nakayama:2015pba} 
for other mechanisms to avoid the above axion isocurvature problem, 
though they did not mention the origin of the PQ symmetry. 
}
We introduce a bulk scalar field 
that is responsible for the stabilization of the size of the extra dimension by the Goldberger-Wise (GW) mechanism~\cite{Goldberger:1999uk}. 
Since the bulk scalar field is coupled to inflaton on our brane, 
the size of extra dimension changes after inflation. 
When the size of extra dimension during inflation is much smaller than the one at present, 
the PQ symmetry can be drastically broken during inflation. 
As a result, PQ breaking operators give the axion a large mass during inflation 
and suppress undesirable axion isocurvature fluctuations. 
After inflation ends, 
the size of the extra dimension becomes large due to an interaction between inflaton and the bulk scalar field, 
and PQ breaking operators are exponentially suppressed. 
In this scenario, the extra dimension gives us mechanisms to explain the origin of the PQ symmetry 
and to avoid the axion isocurvature problem. 
Since the Planck scale is proportional to the size of the extra dimension, 
it also changes at the end of inflation. 
In our scenario, 
dimensionful parameters in the inflaton sector should be rescaled by the ratio of effective Planck scales during and after inflation.

%%%%%%%%%%%%%%%%%%%%%%%%%%%%%%%%%%%%%%%%%%%%%%%%%%%%%%%%%%%%%%%%
\section{extra dimension 
\label{sec:model}}
%%%%%%%%%%%%%%%%%%%%%%%%%%%%%%%%%%%%%%%%%%%%%%%%%%%%%%%%%%%%%%%%

Let us consider a 5D spacetime 
where the 5D manifold is $R^4 \times S / Z_2$ just like the RS I model~\cite{Randall:1999ee} 
and the metric is written as 
\beq
 \dd s^2 = \tilde{n}^2 (t, y) \dd t^2 
 - \tilde{a}^2 (t, y) \dd x_i^2 
 - \tilde{b}^2 (t, y) \dd y^2. 
\eeq
Here, the coordinate of the extra dimension $y$ extends from $-1$ to $1$ 
and the $Z_2$ orbifold symmetry is described by $y \to -y$. 
Two branes are located at the fixed points in the $S^1 / Z_2$ orbifold, 
i.e., $y=0$ and $y=1$.

We introduce matter on branes and a bulk scalar field $\Phi$ 
that is responsible for the stabilization of the size of the extra dimension by the GW mechanism~\cite{Goldberger:1999uk}. 
The action is given by 
\beq
 S &=& \int \dd^5 x \sqrt{g_5} \lmk 
 - \frac{1}{2 \kappa^2} R - \Lambda 
 + \frac{1}{2} \del_\mu \Phi \del^\mu \Phi - \frac{m_\Phi^2}{2} \Phi^2 \rmk \nonumber\\ 
 &+& \int_{y=0} \dd^4 x \sqrt{-g_4} \lmk \mathcal{L}_{\rm m,0} - V_0 (\Phi) \rmk \nonumber\\
 &+& \int_{y=1} \dd^4 x \sqrt{-g_4} \lmk \mathcal{L}_{\rm m,1} - V_1 (\Phi)  \rmk, 
\eeq
where $g_5$ and $g_4$ are the determinants of 5D and 4D metric, respectively. 
The constant $\kappa^2$ is related to the 5D Planck scale $M_5$ as $\kappa^2 = M_5^{-3}$. 
Hereafter, we rewrite the negative bulk cosmological constant $\Lambda$ as $\Lambda = - 6 m_0^2 / \kappa^2$. 
$\mathcal{L}_{\rm m,0}$ and $\mathcal{L}_{\rm m,1}$ 
represent the Lagrangians for matter, 
and the energy momentum tensor corresponding to these Lagrangians is described by 
\beq
 \left. T^\nu_\mu \right\vert_{\rm matter} 
 &=& 
 \frac{1}{\tilde{b}} \delta \lmk y \rmk \diag \lmk \rho_*, 
 - p_*, - p_*, - p_*, 0 \rmk \nonumber\\ 
 &+& 
 \frac{1}{\tilde{b}} \delta \lmk (y-1) \rmk \diag \lmk  \rho, 
 - p, - p, - p, 0 \rmk. 
\eeq
Hereafter, we neglect matter on the brane located at $y=0$ and take $\rho_* = p_* = 0$.

In this letter, 
we take a limit of $m_0 \tilde{b} \ll 1$, 
which implies that the extra dimension is almost flat. 
In this case, the solution of the Einstein equation up to the first order of $\rho$ is given as~\cite{Csaki:1999mp, Cline:2000tx} 
\beq
 \tilde{n} (t,y)&=& 
 1 - \frac{2 + 3 \omega }{12} \kappa^2 b \rho y^2 \\
 \tilde{a} (t,y)&=& a (t) 
 \lmk 1+ \frac{1}{12} \kappa^2 b \rho y^2 \rmk \\
 \tilde{b} (t,y)&=& b, 
\eeq
for $0 \le y \le 1$, where $p = \omega \rho$. 
Throughout this section, 
we consider the case that the radion $b$ stays at its potential minimum [see Eq.~(\ref{b})]. 
The scale factor $a (t)$ obeys the Friedmann equations: 
\beq
 \lmk \frac{\dot{a}}{a} \rmk^2 &\simeq& 
 \frac{8 \pi G}{3} \rho \\
 \lmk \frac{\dot{a}}{a} \rmk^2 - \frac{\ddot{a}}{a} 
 &\simeq& 4 \pi G \lmk \rho + p \rmk \\
 8 \pi G &\simeq& \frac{\kappa^2}{2 b}, 
 \label{M_pl}
\eeq
where $\dot{a} \equiv \dd a / \dd t$. 
Here, a brane tension on a brane has been taken 
so that the cosmological constant is (almost) vanishing in the Friedmann equations. 
Assuming that the VEVs of the bulk scalar field $\Phi$ on our brane and the other brane 
are fixed at $v_1$ and $v_0$ by $V_1 (\Phi)$ and $V_0 (\Phi)$, respectively,%
\footnote{
To justify the flat extra dimension, 
the backreaction of the bulk scalar field on the metric should be subdominant. 
This can be achieved when $v_0, v_1 \ll \kappa^{-1}$~\cite{DeWolfe:1999cp}. 
}
we obtain the static solution of $\Phi$ as 
\beq
 \Phi (y) &=& A e^{- m_\Phi b y} + B e^{- m_\Phi b (1-y)} \\
 A &=& \frac{v_0 - e^{-m_\Phi b} v_1}{1 - e^{-2m_\Phi b}} \\ 
 B &=& \frac{v_1 - e^{-m_\Phi b} v_0}{1 - e^{-2m_\Phi b}}, 
\eeq
for $0 \le y \le 1$. 
Note that $\Phi$, $v_0$, and $v_1$ have mass dimension $3/2$. 
Substituting this solution into the Lagrangian, we obtain the radion potential: 
\beq
 V_r (b) = 
 m_\Phi  \lmk 1- e^{-2 m_\Phi b} \rmk \lmk A^2 + B^2 \rmk. 
\eeq
This implies that the radion stays at 
\beq
 b 
 \simeq 
 \frac{1}{m_\Phi} \ln \frac{v_1}{v_0} 
 ~~~~\text{for}~~ v_1 \ge v_0, 
\label{b}
\eeq
and the size of the extra dimension $b$ is stabilized.

%%%%%%%%%%%%%%%%%%%%%%%%%%%%%%%%%%%%%%%%%%%%%%%%%%%%%%%%%%%%%%%%
\section{origin of PQ symmetry
\label{sec:PQ symmetry}}
%%%%%%%%%%%%%%%%%%%%%%%%%%%%%%%%%%%%%%%%%%%%%%%%%%%%%%%%%%%%%%%%

In this section, we briefly review 
the mechanism to explain the origin of the PQ symmetry in the extra dimension following Refs.~\cite{Izawa:2002qk, Izawa:2004bi}. 
We introduce a bulk hypercolor gauge group $SU(N)_H$ and $N_F$ ($ \ge 4$) pairs of fields $Q^i$ and $\bar{Q}^i$ 
which transforms under the fundamental and anti-fundamental representations of $SU(N)_H$, respectively. 
The first three $Q^i$ and $\bar{Q}^i$ of ``flavour'' indices $i=1,2,3$ 
transform under the fundamental and anti-fundamental representation of $SU(3)_c$, respectively. 
The charge assignment is summarized in Table~\ref{table1}. 
The QCD gauge field is assumed to propagate in the bulk 
while the field $Q$ ($\bar{Q}$) lives on our (the other) brane.%
\footnote{
Gauge anomalies on the branes can be cancelled by a Chern-Simons term in the bulk~\cite{Izawa:2002qk}. 
}
In this setup, 
there is an approximate axial symmetry because interactions between the chiral extra-quarks are suppressed exponentially. 
For example, the mass term of 
\beq
 M_Q Q \bar{Q} + {\rm H.c.}, 
 \label{M_Q} 
\eeq
is suppressed as $M_Q \sim b^{-1} e^{- M_5 b }$.

%%%%%%%%%%%%%%%%%%%%%%%%%%%%%%%%%%%%%%%%%%%%%%%%%%%%%%%%%%%%%%%%
\begin{table}\begin{center}
{\renewcommand\arraystretch{1.5}
\begin{tabular}{|c|p{1.4cm}|p{1.4cm}|p{1.4cm}|p{1.4cm}|}
  \hline
    & \hfil $Q^{i(=1,2,3)}_a$ \hfil & \hfil $\bar{Q}^{i(=1,2,3)}_a$ \hfil & \hfil $Q^{i (\ge 4)}_a$ \hfil & \hfil $\bar{Q}^{i (\ge 4)}_a$ \hfil \\
  \hline
  \hfil $SU(3)_c$ \hfil & \hfil {\bf 3} \hfil & \hfil ${\bf 3}^*$ \hfil & \hfil {\bf 1} \hfil & \hfil {\bf 1} \hfil \\
  \hline
  \hfil $SU(N)_{\rm H}$ \hfil & \hfil {\bf N} \hfil & \hfil ${\bf N}^*$ \hfil & \hfil {\bf N} \hfil & \hfil ${\bf N}^*$ \hfil \\
\hline
\end{tabular}
}
\end{center}
\caption{Charge assignment for matter fields.
\label{table1}}
\end{table}
%%%%%%%%%%%%%%%%%%%%%%%%%%%%%%%%%%%%%%%%%%%%%%%%%%%%%%%%%%%%%%%%

We assume that the hypercolor gauge interaction confines at a scale $f_a$ and a chiral condensate develops as 
\beq
 \la Q^i \bar{Q}^j \ra \simeq f_a^3 \delta^{ij}. 
\eeq
In terms of the $SU(N)_H$ gauge theory, 
the $U(N_F)_L \times U(N_F)_R$ flavour symmetry is spontaneously broken to the diagonal $U(N_F)_V$ symmetry 
and there would be a large number of NG bosons in the low energy effective theory. 
However, 
the flavour symmetry is explicitly broken by $SU(3)_c$ gauge interactions 
and $SU(3)_c$ charged NG bosons acquire masses via $SU(3)_c$ radiative corrections. 
In addition, 
there is $U(1) \lkk SU(N)_H \rkk^2$ anomaly, 
so that one linear combination of $SU(3)_c$ singlet NG bosons 
obtains a mass through nonperturbative effects. 
Thus, 
there are $(N_F - 3)^2$ NG bosons in the effective theory below the energy scale of $f_a$. 
One of these NG bosons that is associated with the following $U(1)$ symmetry can be identified with axion: 
\beq
\left\{
\bea{ll}
 Q^{i} (\bar{Q}^{i}) \to e^{i \alpha/3} Q^{i}  (\bar{Q}^{i})
 &~~\text{for}~~ i = 1,2,3 \\
 Q^{i}  (\bar{Q}^{i}) \to e^{-i \alpha/(N_F - 3)} Q^{i} (\bar{Q}^{i})
 &~~\text{for}~~ i \ge 4. 
\eea
\right.
\eeq
This is actually free from $U(1) \lkk SU(N)_H \rkk^2$ anomaly. 
Hereafter we denote this $U(1)$ symmetry as $U(1)_{\rm PQ}$.

The $U(1)_{\rm PQ}$ symmetry has $U(1)_{\rm PQ} \lkk SU(3)_c \rkk^2$ anomaly, 
so that the axion acquires a mass through nonperturbative effects 
after the QCD phase transition~\cite{'tHooft:1976up, 'tHooft:1976fv}. 
When the Hubble parameter becomes comparable to the axion mass, 
the axion starts to oscillate around its minimum~\cite{Preskill:1982cy, Abbott:1982af, Dine:1982ah}. 
The present energy density of the axion oscillation is calculated as~\cite{Wantz:2009it, Kawasaki:2014sqa}
\beq
 \Omega_a h^2 
 \simeq 
 0.011 \theta_{\rm ini}^2 
 \lmk \frac{f_a / N_{\rm DW}}{10^{11} \GEV} \rmk^{1.19} 
  \lmk \frac{\Lqcd}{400 \MEV} \rmk, 
  \label{axion abundance}
\eeq
where $h$ is the Hubble parameter in units of $100 \ {\rm km/s/Mpc}$ 
and $\theta_{\rm ini}$ is the initial misalignment angle. 
The domain wall number $N_{\rm DW}$ is equal to $N$ in our model 
because there are $N$ flavours in terms of the $SU(3)_c$ symmetry. 
The axion abundance is consistent with the observed DM abundance $\Omega_{\rm DM}^{\rm obs} h^2 \simeq 0.12$ 
when the axion decay constant is given as 
\beq
 f_a \simeq 7 \times 10^{11} \GEV \times N_{\rm DW} \theta_{\rm ini}^{-1.68}, 
 \label{F_a from DM}
\eeq
where we use $\Lqcd = 400 \MEV$. 
Note that such an intermediate scale can be naturally realized in our model 
because the PQ breaking scale $f_a$ is determined by the strong dynamics of the $SU(N)_H$ gauge theory.

Let us check whether or not PQ breaking operators are sufficiently suppressed to explain the smallness of the strong CP phase. 
The operator of Eq.~(\ref{M_Q}) gives the axion a mass of order $\sqrt{M_Q f_a}$ 
at a minimum that is generally different from the one determined by the original vacuum angle $\theta_0$. 
Thus, it may induce a shift in the strong CP phase as 
\beq
 \Delta \theta \sim \lmk \frac{\sqrt{M_Q f_a}}{m_a} \rmk^2, 
\eeq
where $m_a$ is the axion mass at the low energy: 
\beq
 m_a \simeq \frac{m_u m_d}{\lmk m_u + m_d \rmk^2} \frac{m_\pi f_\pi}{f_a / N_{\rm DW}}. 
\eeq
Here, $f_\pi$ is the pion decay constant, 
and $m_u$, $m_d$, and $m_\pi$ are the masses of up quark, down quark, and pion, respectively. 
Requiring $\Delta \theta < \theta_{\rm obs} \sim 10^{-10}$ and using $M_Q \sim b^{-1} e^{-M_5 b}$, 
we find that $M_5 b \gtrsim 150 + 3 \ln ( f_a / 10^{12} \GEV)$ is sufficient to solve the strong CP problem.

However, operators involving either $Q$ or $\bar{Q}$ are expected to be suppressed only by powers of $M_5^{-1}$ on each brane. 
In the case of $N = 3$, for example, we can write the following terms on the branes: 
\beq
 &&\int_{y = 0} \dd^4 x \sqrt{-g_4} \frac{y_Q}{M_5^5} \lmk Q Q Q \rmk^2 \nonumber \\
 &&\quad + 
 \int_{y = 1} \dd^4 x \sqrt{-g_4} \frac{y_{\bar{Q}}}{M_5^5} \lmk \bar{Q} \bar{Q} \bar{Q} \rmk^2 
 + {\rm H. c.}, 
 \label{PQ breaking terms}
\eeq
where $y_Q$ and $y_{\bar{Q}}$ are coupling constants. 
These terms also give a mass to the axion and may shift the strong CP phase as 
\beq
 \Delta \theta \sim 5 \times 10^{-13} y_Q y_{\bar{Q}} 
 \lmk \frac{f_a}{10^{12} \GEV} \rmk^{14} 
 \lmk \frac{M_5}{\Mpl} \rmk^{-10}. 
\eeq
Noting $M_5 = \Mpl / \sqrt{2 b M_5} \lesssim \Mpl / 18$, 
this is marginally consistent with the present upper bound on the strong CP phase 
for $y_Q y_{\bar{Q}} \lesssim 10^{-3}$. 
Although the bulk scalar field $\Phi$ and radion $b$ can interact with $Q$ and $\bar{Q}$ on each brane, 
these contributions are also suppressed in the same way. 
Note that $\Delta \theta$ is much more suppressed when we consider $N \ge 5$. 
In the case of $N = 5$ and $N_F \ge 5$, for example, 
the lowest dimension operator is written as $(QQQQQ)^2 / M_5^{11}$ 
and the shift in the strong CP phase is at most of order $10^{-50}$ for $f_a = 3 \times 10^{12} \GEV$. 
Note that we have assumed that light colored particles other than $Q$ and $\bar{Q}$ are absent in the bulk and on the branes. 
This ensures that terms like Eq.~(\ref{PQ breaking terms}) are the lowest dimension operators, 
so that explicit breakings of the PQ symmetry is sufficiently suppressed to solve the strong CP problem.

%%%%%%%%%%%%%%%%%%%%%%%%%%%%%%%%%%%%%%%%%%%%%%%%%%%%%%%%%%%%%%%%
\section{suppressing isocurvature fluctuations 
\label{sec:isocurvature}}
%%%%%%%%%%%%%%%%%%%%%%%%%%%%%%%%%%%%%%%%%%%%%%%%%%%%%%%%%%%%%%%%

When the PQ symmetry is broken during inflation and the mass of axion is much less than the Hubble parameter, 
the axion acquires sizable quantum fluctuations 
and predicts isocurvature perturbations which can be observed in CMB fluctuations~\cite{Axenides:1983hj, Seckel:1985tj, Turner:1990uz}. 
However, the observation of CMB fluctuations reveals that they are predominantly adiabatic 
and puts an upper bound on the amount of isocurvature perturbations. 
This upper bound implies that the energy scale of inflation has to be smaller than of order $10^7 \GEV$~\cite{Ade:2015lrj}. 
Such a small energy scale severely restricts inflation models. 
This is called the axion isocurvature problem.

In our model, the mass of axion is suppressed exponentially as $e^{- M_5 b}$ (see Eq.~(\ref{M_Q})). 
This implies that 
if the size of the extra dimension $b$ during inflation is smaller than its present value by more than one order of magnitude, 
the mass of axion can be larger than the Hubble parameter during inflation and axion fluctuations can be suppressed. 
This scenario can be realized in the following way. 
Let us consider a scalar field $\cphi$ (such as an inflaton or a waterfall field) which lives on our brane 
and whose VEV changes from $0$ to $\cphi_0$ after inflation. 
We assume that $\cphi$ and the bulk scalar field $\Phi$ have an interaction term such as 
\beq
 V_1 (\Phi) = \frac{\lambda_1}{4 M_5^2} ( \Phi^2 - v_1^2)^2 - \frac{\lambda_{\rm int}}{4 M_5} \Phi^2 \cphi^2 + \Lambda_1, 
 \label{V_1}
\eeq
where $\lambda_1$ and $\lambda_{\rm int}$ are dimensionless parameters. 
Then the VEV of $\Phi$ on our brane changes from $v_1$ to 
$v'_1 \equiv  \sqrt{v_1^2 + (\lambda_{\rm int} / 2 \lambda_1) M_5 \cphi_0^2}$ after inflation, 
which means that the size of the extra dimension changes after inflation via Eq.~(\ref{b}). 
As a result, the mass of axion can be much larger than its present value 
and can be larger than the Hubble parameter during inflation. 
For example, when $M_5 b^{\rm inf} \simeq 10$ during inflation, 
the axion mass is as large as $m_a \sim \sqrt{M_Q f_a} \sim 10^{12} \GEV$.

In order to solve the strong CP problem, 
the axion mass has to be suppressed and the size of the extra dimension should satisfy 
$M_5 b^{\rm now} \gtrsim 150 + 3 \ln (f_a / 10^{12} \GEV)$ at the present epoch. 
In the case of $v_1 / v_2 = 1.1$, $v'_1 / v_2 = 5$, and $m_\Phi / M_5 = 0.01$, for instance, 
we obtain desired values: $M_5 b^{\rm inf} \simeq 9.5$ during inflation 
and $M_5 b^{\rm now} \simeq 161$ at present. 
Note that the axion VEV during inflation is generally different from the one at the present epoch. 
This is because the former one is determined by the phase of the mass term of Eq.~(\ref{M_Q}) 
while the latter one is determined by the original vacuum angle $\theta_0$. 
Therefore the axion starts to oscillate around the latter minimum at the QCD phase transition 
and its abundance is given by Eq.~(\ref{axion abundance}).

%%%%%%%%%%%%%%%%%%%%%%%%%%%%%%%%%%%%%%%%%%%%%%%%%%%%%%%%%%%%%%%%
\section{predictions
%inflation parameters
\label{sec:predictions}}
%%%%%%%%%%%%%%%%%%%%%%%%%%%%%%%%%%%%%%%%%%%%%%%%%%%%%%%%%%%%%%%%

\subsection{inflation}
\label{sec:inflation}

In our scenario, the extra dimension is almost flat and its size is of order the 5D Planck length $M_5^{-1}$ 
so that high scale inflation can be realized. 
Although such a small extra dimension cannot be directly accessible at collider experiments, 
predictions of CMB fluctuations may be modified in our scenario. 
This is because the Planck scale during inflation $\Mpl^{\rm inf}$ is different from the one in the present epoch $\Mpl^{\rm now}$ 
as $\Mpl^{\rm inf} = \delta^{1/2} \Mpl^{\rm now}$ (see Eq.~(\ref{M_pl})), 
where $\delta = b^{\rm inf} / b^{\rm now}$ $(= O(10^{-1}))$.

For simplicity, we consider the case that 
the size of the extra dimension is constant during the time that the CMB modes exit the horizon. 
Since curvature perturbations are conserved outside the horizon, 
the resulting CMB spectrum is easily calculated by rescaling the Planck scale. 
During inflation, the Hubble parameter is given by~\cite{Kaiser:1994vs}
\beq
 H^2 \equiv \lmk \frac{1}{a} \frac{\dd a}{\dd t} \rmk^2 \simeq \frac{1}{3 \delta \Mpl^2} V. 
\eeq
The spectral index is calculated from $n_s = 1 - 6 \epsilon + 2 \eta$, 
where the slow roll parameters are given by 
\beq
 \epsilon = \frac{\delta \Mpl^2}{2} \lmk \frac{V'}{V} \rmk^2 
 \\
 \eta = \delta \Mpl^2 \lmk \frac{V''}{V} \rmk, 
\eeq
where the prime denotes a derivative with respect to inflaton. 
These results imply that slow roll inflation requires a steeper potential than ordinary inflation models. 
The amplitude of scalar perturbations of the metric is given by 
\beq
 \Delta_{\mathcal{R}}^2 \simeq \frac{V /  \epsilon }{\delta^2 \Mpl^4 24 \pi^2}, 
\eeq
and the tensor-to-scalar ratio is given by $r = 16 \epsilon$. 
It is clear that 
once all dimensionful parameters in the inflaton sector are rescaled by a factor of $\delta^{1/2}$ 
(e.g., $\cphi_{\rm inf} \to \delta^{1/2} \cphi_{\rm inf}$ and $V \to \delta^2 V$), 
we reproduce standard predictions of inflation. 
This is because the rescale of the Planck scale can be performed by a conformal transformation, 
which does not change the values of the dimensionless parameters $\epsilon$, $\eta$, and $\Delta_{\mathcal{R}}$.

We assume that the VEV of $\cphi$ in Eq.~(\ref{V_1}) changes from $0$ to $\cphi_0$ after inflation. 
This is naturally realized when we consider a hybrid inflation model and identify $\cphi$ as a waterfall field. 
Denoting inflaton as $I$, we write the potential in the inflaton sector as 
\beq
 V_{\rm inf} = \frac{\kappa}{4} \cphi^2 I^2 + \frac{\lambda_{\cphi}}{4} \cphi^4 + V_I (I). 
\eeq
The inflaton $I$ has a large VEV during inflation 
and slowly rolls toward the critical point $I_c \equiv  ( \lambda_{\rm int} v_1^2 / \kappa M_5 )^{1/2}$ due to an inflaton potential $V_I (I)$. 
When the VEV of the inflaton decreases to $I_c$, 
the waterfall field $\cphi$ starts to oscillate around the true minimum and the inflation ends. 
At the same time, the VEV of the bulk scalar field $\Phi$ on our brane changes from $v_1$ to $v'_1$ 
and the size of the extra dimension changes from $b^{\rm inf}$ to $b^{\rm now}$.

Finally, we should note that 
the oscillations of $\Phi$ and the radion $b$ may be induced after inflation due to the change of their potential minimum. 
When we introduce an interaction of $\Phi N_R N_R$ on our brane, where $N_R$ is the right-handed neutrino, 
$\Phi$ can decay into right-handed neutrinos and then the right-handed neutrinos decay into standard-model particles. 
Since the radion $b$ couples with SM particles via Planck-suppressed operators, 
it can also decay into radiation. 
In our model, the mass of the radion is given by 
\beq
 m_b^2 = \frac{2}{3 \Mpl^2} (b^{\rm now})^2 m_\Phi^3 \frac{v_1^2 v_2^2}{\abs{v_1^2 - v_2^2}}, 
\eeq
after its kinetic term is canonically normalized. 
This is of order $10^{13} \GEV$ for $v_1, v_2 = \kappa^{-1}/100$, 
so that the reheating temperature can be as large as $10^{10} \GEV$ 
even if they dominate the Universe. 
Such a high reheating temperature is consistent with the realization of the thermal leptogenesis~\cite{Fukugita:1986hr}.

\subsection{dark radiation}
\label{sec:DR}

In our model, 
there are $(N_F - 3)^2 - 1$ massless NG bosons in addition to the axion, 
so that they may contribute to the energy density of the Universe as dark radiation~\cite{Turner:1986tb, Nakayama:2010vs, Weinberg:2013kea}. 
Since they interact with the SM particles via interactions suppressed by some powers of $f_a$, 
their decoupling temperature is roughly an order of magnitude below $f_a$ (see e.g., Refs.~\cite{Masso:2002np, Kawasaki:2015ofa}). 
Therefore if 
the maximal temperature after inflation is lower than $f_a$ 
and 
the reheating temperature can be as large as of order $f_a/10$, 
the NG bosons may be thermalized after inflation without restoring the PQ symmetry. 
In fact, such a high reheating temperature is favoured to realize the thermal leptogenesis~\cite{Fukugita:1986hr}. 
Once the NG bosons and axion are thermalized and decoupled at a high temperature, 
they contribute to the energy density of the Universe as dark radiation. 
Its amount is conventionally expressed by the effective neutrino number 
and the result is calculated as~\cite{Kawasaki:2015ofa} 
\beq 
 N_{\rm eff} \simeq N_{\rm eff}^{({\rm SM})} + 0.027 \times (N_F - 3)^2, 
\eeq
where $N_{\rm eff}^{({\rm SM})}$ ($\simeq 3.046$) is the SM prediction. 
The present constraint on $N_{\rm eff}$ is $N_{\rm eff} = 2.99 \pm 0.39$ $(95\% \ {\rm C.L.})$~\cite{Aver:2013wba, Planck:2015xua}, 
which implies that $N_F$ should be smaller than or equal to $6$. 
The ground-based Stage-IV CMB polarization experiment CMB-S4 
will measure the effective neutrino number with a precision of 
$\Delta N_{\rm eff} = 0.0156$ within one sigma level~\cite{Wu:2014hta} (see also Ref.~\cite{Abazajian:2013oma}). 
The number of flavours can be measured indirectly via the observation of the amount of dark radiation.%
\footnote{
Although the amount of dark radiation for $N_F = 6$ is exactly the same with the one predicted in Ref.~\cite{Kawasaki:2015ofa}, 
we could distinguish between these models via the observation of an exotic kaon decay process. 
}

%%%%%%%%%%%%%%%%%%%%%%%%%%%%%%%%%%%%%%%%%%%%%%%%%%%%%%%%%%%%%%%%
\section{discussion and conclusions
\label{sec:conclusion}}
%%%%%%%%%%%%%%%%%%%%%%%%%%%%%%%%%%%%%%%%%%%%%%%%%%%%%%%%%%%%%%%%

We have proposed a QCD axion model in a 5D spacetime 
to explain the origin of the PQ symmetry and avoid the axion isocurvature problem. 
Each extra quark-antiquark pair lives on branes separately 
so that they have an accidental PQ symmetry~\cite{Izawa:2002qk, Izawa:2004bi}. 
We assume an interaction between inflaton 
and a bulk scalar field that is responsible for the stabilization of the size of the extra dimension. 
As a result, the size of the extra dimension during inflation can be different from the one at present 
and PQ breaking operators can be efficient during inflation. 
We have shown that the axion mass from PQ breaking operators can be larger than the Hubble parameter during inflation 
to suppress axion isocurvature fluctuations 
while the PQ symmetry is sufficiently preserved after inflation to solve the strong CP problem. 
Therefore, the extra dimension explains not only the origin of the PQ symmetry 
but also suppression of axion isocurvature fluctuations.

Since the size of the extra dimension changes after inflation, 
all dimensionful parameters in the inflaton sector should be rescaled by a certain factor compared with the ones in ordinary inflation models. 
In addition, our model may predict a sizable amount of dark radiation which would be detected in the near future. 
Finally, we comment on our assumption that the bulk scalar $\Phi$ and the ration $b$ stay at their minimum. 
This is actually justified during and well after inflation because of the redshift effect. 
The dynamics of $\Phi$ and $b$ just after inflation might be complicated and will be discussed elsewhere.

\vspace{1cm}

%---------------SECTION------------------%
%
\section*{Acknowledgments}
This work is supported by Grant-in-Aid for Scientific Research 
from the Ministry of Education, Science, Sports, and Culture
(MEXT), Japan, 
No. 25400248 (M.K.), No. 26104009, and No. 26287039 (T.T.Y), 
World Premier International Research Center Initiative
(WPI Initiative), MEXT, Japan,
and the Program for the Leading Graduate Schools, MEXT, Japan (M.Y.).
M.Y. acknowledges the support by the JSPS Research Fellowships for Young Scientists, No. 25.8715.
%
%---------------SECTION------------------%

\vspace{1cm}

%%%%%%%%%%%%%%%%%%%%%%%%%%%%%%%%%%%%

%%%%%%%%%%%%%%%%%%%%%%%%%%%%%%%%%%%%

\end{document}